\documentclass[
    ,final            
  ]
  {aipproc}

\layoutstyle{6x9}

\begin{document}

\title{On the relation between $E(5)-$models and the interacting boson
  model}


\author{Jos\'e Enrique Garc\'{\i}a-Ramos}{
  address={Departamento de
  F\'{\i}sica Aplicada, Universidad de Huelva, 21071 Huelva, Spain},
}

\author{Jos\'e M. Arias}{
  address={Departamento de F\'{\i}sica At\'omica, Molecular y
  Nuclear, Facultad de F\'{\i}sica, Universidad de Sevilla,
  Apartado~1065, 41080 Sevilla, Spain}
}

\begin{abstract}
  The connections between the $E(5)-$models (the original $E(5)$ using
  an infinite square well, $E(5)-\beta^4$, $E(5)-\beta^6$ and
  $E(5)-\beta^8$), based on particular solutions of the geometrical
  Bohr Hamiltonian with $\gamma$-unstable potentials, and the
  interacting boson model (IBM) are explored. For that purpose, the
  general IBM Hamiltonian for the $U(5)-O(6)$ transition line is used
  and a numerical fit to the different $E(5)-$models energies is
  performed. It is shown that within the IBM one can
  reproduce very well all these $E(5)-$models. The agreement is the
  best for $E(5)-\beta^4$ and reduces when passing through
  $E(5)-\beta^6$, $E(5)-\beta^8$ and $E(5)$, where the worst agreement
  is obtained (although still very good for a restricted set of lowest
  lying states).  The fitted IBM Hamiltonians correspond to energy
  surfaces close to those expected for the critical point.
\end{abstract}

\maketitle


\section{Introduction}
\label{sec-intro}
Both, the Bohr-Mottelson (BM) collective model \cite{bm} and the
interacting boson model (IBM) \cite{ibm} have thoroughly been used to
study the same kind of nuclear structure problems.  Although very
different in their formulation, the two models present clear
relationships.  Both models have three particular cases that can be
easily solved and for which a clear correspondence can be done: i)
spherical nucleus, ii) $\gamma$-unstable deformed rotor and, iii)
axial rotor.  For transitional situations and, specially in the phase
transition areas, the correspondence between the two models is
difficult \cite{Rowe05}.  This suggests, for the case of transitional
Hamiltonians, to look for the connection between BM and IBM through
numerical studies.

In this work and in Ref.~\cite{Garc08}, 
we concentrate on $E(5)$ and related models: the
original $E(5)$ (infinite square well potential) \cite{Iach00} and, 
$E(5)$ with a
potential $\beta^4, \beta^6$ and, $\beta^8$, respectively \cite{Bona04}. All these
models are produced in the BM scheme and a natural question is to ask
for the corresponding equivalence in the IBM. Is the IBM able for
producing the same spectra and transition rates? If yes, does the IBM
Hamiltonian correspond to a critical point? This work is intended to
answer these questions for those models and analyze the convergence as
a function of the boson number.
This procedure will
allow to establish the IBM Hamiltonian which best fit the different
$E(5)-$models and their relation with the critical points.

\section{The IBM fit to $E(5)-$models}
\label{sec-fit}

The most general, including up to two-body terms, IBM Hamiltonian can
be written in multipolar form as,
\begin{eqnarray}
  \label{ham1}
  \hat H&=&\varepsilon_d \hat n_d +
  \kappa_0 \hat P^\dag \hat P
  +\kappa_1 \hat L\cdot \hat L+
  \kappa_2 \hat Q \cdot \hat Q+
  \kappa_3 \hat T_3\cdot\hat T_3 +\kappa_4 \hat T_4\cdot\hat T_4
\end{eqnarray}  
where the definition of the different operators can be found in
Ref.~\cite{Fran94}.

The $E(5)-$models are intended to be of use for $\gamma$-unstable
nuclei having $O(5)$ as symmetry algebra.
For the construction of an IBM $\gamma$-unstable transitional
Hamiltonian it is sufficient to impose in Eq.~(\ref{ham1})
$\kappa_2=0$. 
If additionally, we want to construct an IBM transitional Hamiltonian
that preserves the $O(5)$ symmetry
we have to impose the constraint
$\kappa_1-\kappa_3/10-\kappa_4/14=0$ \cite{Garc08}.  
In practice, 
we 
do not
impose the later restriction
but, as it will be
shown, this condition will be fulfilled in every fit. 
It is worth noting that in Ref.~\cite{Garc08} we used the extra
constraint $\kappa_4=0$ for simplicity and, the raised conclusions are
qualitatively identical to the ones obtained in the present
contribution.  
\begin{figure}[hbt]
  \centering
  \includegraphics[width=7.cm]{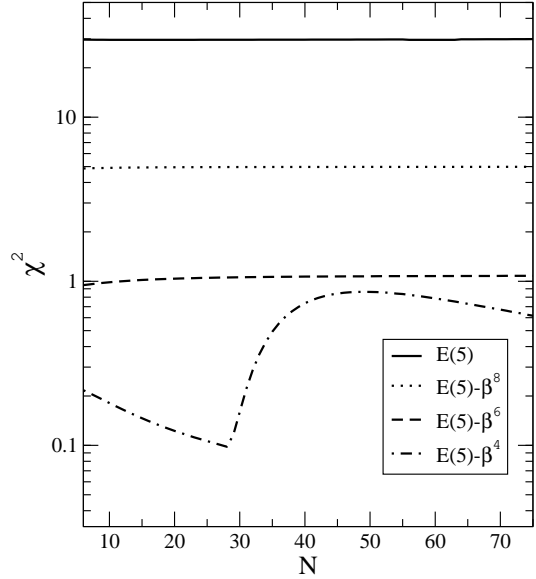}
  \caption{$\chi^2$ for the IBM fit to the energy
    levels of the different $E(5)$-models, as a function of $N$.}
  \label{fig-chi2-e5-t4t4}
\end{figure}
    
In order to perform the fit, we minimize a standard $\chi^2$ function for the
energies, using
$\varepsilon_d$, $\kappa_0$, $\kappa_1$, $\kappa_3$, and $\kappa_4$ as free
parameters and $\kappa_2$ fixed to zero.
We have done fits of the IBM Hamiltonian (\ref{ham1}) parameters, as a
function of $N$, so
as to reproduce as well as possible the energies 
generated by the different $E(5)-$models (see Ref.~\cite{Garc08} for
more details about the fitting procedure). The value 
of the $\chi^2$ for a best fit to
the different $E(5)-$models as a function of $N$ is shown in 
Fig.~\ref{fig-chi2-e5-t4t4}.
It is clearly
observed that for any $N$ the agreement between the fitted IBM and the
$E(5)-\beta^4$ model is excellent and is getting worse for
$E(5)-\beta^6$, $E(5)-\beta^8$, up to reach $E(5)$ which is the worst
case. In particular $\chi^2(E(5)-\beta^4) \approx \chi^2(E(5))/50$.
It is worth noting that these results change slowly with the boson
number and in all cases the $\chi^2$ value is approximately constant, 
except for $E(5)-\beta^4$ which is decreasing. 
If the calculations are extended till $N=1000$ bosons
(see Ref.~\cite{Garc08}) one observes how $\chi^2$ values will
continue having 
finite values, close to the ones given in figure
\ref{fig-chi2-e5-t4t4}, except for the case $E(5)-\beta^4$ which
decreases and, it is expected to vanish for $N\rightarrow\infty$, as it
was shown in Ref.~\cite{Aria03}.

To have a clearer idea of the degree of agreement between the fitted
IBM results with the data from the $E(5)-$models we analyze the case
of 
$N=60$. 
In Table \ref{table-parameters-t4t4} we give 
the parameters of the Hamiltonian. 
\begin{table}
\begin{tabular}{|c|c|c|c|c|c|}
\hline
              &$\varepsilon_d$ &$\kappa_0$ &$\kappa_1$&$\kappa_3$&$\kappa_4$\\
\hline
$E(5)$        &251.84     &0.16   &23.5570   &-16.6450  &352.83 \\
$E(5)-\beta^8$&1499.20    &27.11  &12.8750   &4.0282    &174.52 \\
$E(5)-\beta^6$&2482.80    &42.66  &4.3049    &10.1250   &46.08  \\
$E(5)-\beta^4$&2543.00    &39.92  &0.7143    &6.2221    &1.29   \\
\hline
\end{tabular}
  \caption{Parameters of the IBM Hamiltonians  
    used in table  \ref{table-states-t4t4}.} 
  \label{table-parameters-t4t4}
\end{table}
Note that the best fit parameters give
rise approximately to the cancellation of the quadratic Casimir
operator for $O(3)$, {\it i.e.}  $\kappa_1-\kappa_3/10-\kappa_4/14=0$. This
condition is approximately fulfilled for any number of bosons.

In Table \ref{table-states-t4t4} we present the value of the
energies for $N=60$.
The
agreement for $E(5)-\beta^4$, $E(5)-\beta^6$, and $E(5)-\beta^8$ is
really remarkable for all the states.  Only in the case of $E(5)$, one
can observe small discrepancies in the $\xi=2$
and $\xi=3$ bands, while for $\xi=1$ the agreement is perfect. 
This impressive one-to-one
correspondence between the IBM and the $E(5)-$ states, at least for
some bands, suggests the existence  of an underlying phenomenon similar to
the quasidynamical symmetry \cite{Rowe05,Rowe04a} which is called
quasi-critical point symmetry \cite{Garc08}.

Once the parameters of the Hamiltonian have been fixed we check the
wave functions through the calculations of the relevant $B(E2)$
values. 
For all the cases, the agreement between the IBM calculations
and the $E(5)-$ counterpart is reasonable \cite{Garc08}.  

Another consequence of the excellent 
agreement between the $E(5)-$models and the IBM is 
that it is impossible to discriminate, from a
experimental point of view, between a $E(5)-$model and its IBM
counterpart.

\begin{table}
\begin{tabular}{|c|c||c|c||c|c||c|c||c|c|}
\hline
 &$\xi,\tau$&E(5)&IBM&E(5)-$\beta^8$&IBM&E(5)-$\beta^6$&IBM&E(5)-$\beta^4$&IBM\\
\hline 
$0_1^+$     &1,0& 0.000& 0.000 & 0.000& 0.000 & 0.000&0.000&0.000&0.000\\
$2_1^+$     &1,1& 1.000& 1.000 & 1.000& 1.000 & 1.000&1.000&1.000&1.000\\
$4_1^+$     &1,2& 2.199& 2.196 & 2.157& 2.156 & 2.135&2.137&2.093&2.092\\
$2_2^+$     &1,2& 2.199& 2.195 & 2.157& 2.156 & 2.135&2.137&2.093&2.092\\
$0_2^+$     &2,0& 3.031& 3.035 & 2.756& 2.757 & 2.619&2.622&2.390&2.389\\
$6_1^+$     &1,3& 3.590& 3.587 & 3.459& 3.457 & 3.391&3.393&3.265&3.264\\
$4_2^+$     &1,3& 3.590& 3.586 & 3.459& 3.457 & 3.391&3.393&3.265&3.264\\
$3_1^+$     &1,3& 3.590& 3.586 & 3.459& 3.457 & 3.391&3.393&3.265&3.264\\
$0_3^+$     &1,3& 3.590& 3.586 & 3.459& 3.456 & 3.391&3.393&3.265&3.264\\
$2_3^+$     &2,1& 4.800& 4.761 & 4.255& 4.235 & 4.012&3.977&3.625&3.632\\
$6_2^+$     &1,4& 5.169& 5.172 & 4.894& 4.896 & 4.757&4.756&4.508&4.508\\
$5_1^+$     &1,4& 5.169& 5.172 & 4.894& 4.895 & 4.757&4.756&4.508&4.508\\
$4_3^+$     &1,4& 5.169& 5.172 & 4.894& 4.895 & 4.757&4.756&4.508&4.508\\
$2_4^+$     &1,4& 5.169& 5.171 & 4.894& 4.895 & 4.757&4.756&4.508&4.508\\
$4_4^+$     &2,2& 6.780& 6.683 & 5.874& 5.843 & 5.499&5.424&4.918&4.935\\
$2_5^+$     &2,2& 6.780& 6.683 & 5.874& 5.843 & 5.499&5.424&4.918&4.935\\
$0_4^+$     &3,0& 7.577& 7.522 & 6.364& 6.372 & 5.887&5.805&5.153&5.176\\
$2_7^+$     &3,1&10.107& 9.974 & 8.269& 8.293 & 7.588&7.448&6.563&6.606\\
\hline
\end{tabular}
  \caption{Comparison of energy levels for fitted IBM Hamiltonians, 
    with $N=60$, compared with those provided by the $E(5)$-models 
    (see text). 
  }
  \label{table-states-t4t4}
\end{table}

\section{The critical Hamiltonian}
\label{sec-crit}
One of the most attractive features of the $E(5)-$models
is that they are supposed to describe, at different
approximation levels, the critical point in the transition from
spherical to deformed $\gamma$-unstable shapes. Since they are connected to a
given IBM Hamiltonian, as shown in the preceding section, this should
correspond to the critical point in the transition from $U(5)$ to
$O(6)$ IBM limits. Is this the case for the
fitted IBM Hamiltonians obtained in the preceding section?  

\begin{figure}[hbt]
  \centering
  \includegraphics[width=7.cm]{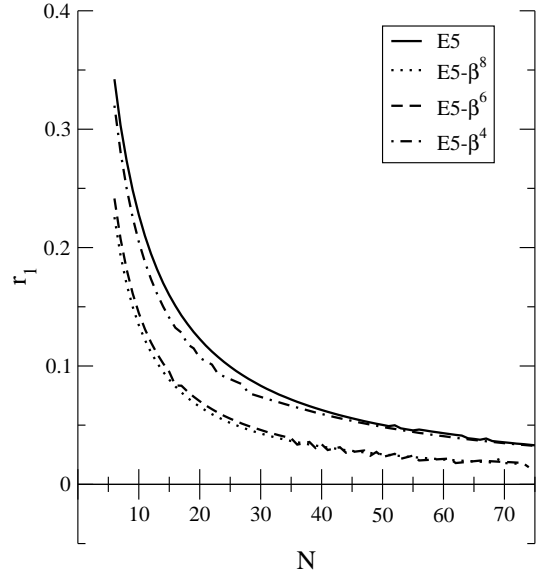}
  \caption{Values of $r_1$ (see text for definition)
    as a function of $N$ for the fitted IBM Hamiltonians.}
  \label{fig-r1-T4T4}
\end{figure}

To analyze critical points and phase transitions in the IBM, 
one of the options is to use the intrinsic state formalism 
\cite{IF}
which introduces the shape variables $(\beta,\gamma)$ in the IBM. 
Due to the characteristics of the Hamiltonian we are working
on, we can only observes second order phase transitions.
To know if we
have  a critical Hamiltonian, it is convenient to use 
the concept of IBM ``essential'' parameters $(r_1,r_2)$ \cite{Lope96}, 
directly related with the
parameters of the Hamiltonian (\ref{ham1}), that allows to quantify the closeness
to a critical point. In particular, in our case $r_2$ always vanishes
(because $k_2=0$) while $r_1$ is defined as,
\begin{equation}
  r_1=\frac{-\kappa_0+(\varepsilon_d+
  6\,\kappa_1 +\frac{7}{5}\,\kappa_3+\frac{9}{5}\,\kappa_4) /(N-1)}
{\kappa_0 +\frac{36}{35}\,\kappa_4+
    (\varepsilon_d+
  6\,\kappa_1 +\frac{7}{5}\,\kappa_3+\frac{9}{5}\,\kappa_3)/(N-1)}. 
  \label{r1}
\end{equation}
In this language, a critical Hamiltonian corresponds to $r_1=0$. In
figure \ref{fig-r1-T4T4} the values of $r_1$ as a function of $N$
for the IBM Hamiltonians obtained from the fit are presented for the
different studied $E(5)-$models.  
In all the cases it is observed an approximation to $r_1=0$ as the
number of bosons increase. 
For the
$E(5)-\beta^4$ model it is known that $r_1=0$ is reached for very
large number of bosons \cite{Aria03}.
 
\section{Conclusions}
\label{sec-conclu}
In this paper, we have studied the connection between the
$E(5)-$models and the IBM on the basis of a numerical mapping between
both models. 
We have shown that it is possible, in all cases, to establish a
one-to-one mapping between the $E(5)-$models and the IBM with a
remarkable agreement for the energies and the $B(E2)$ values. 
Globally, the best agreement is obtained for the
$E(5)-\beta^4$ Hamiltonian and the worst for the $E(5)$ case.
All this suggests the presence of an underlying
quasi-critical point symmetry \cite{Garc08}.

Another consequence of this excellent
agreement is that it is impossible, from a experimental point of view,
to discriminate between a $E(5)$-model and its corresponding IBM
Hamiltonian when only few low-lying states are considered.

We have also proved that all the $E(5)-$models correspond to IBM
Hamiltonians very close to the critical area, $|r_1|<0.05$.
Therefore, one can say that the $E(5)-$models are appropriated to
describe transitional $\gamma-$unstable regions close to the critical
point.

\begin{theacknowledgments}
This work has been partially supported
by the Spanish MEC and by the European FEDER under 
projects number FIS2005-01105,
FPA2006-13807-C02-02 and FPA2007-63074, and by the Junta de
Analuc\'{\i}a under projects FQM160, FQM318, P05-FQM437 and
P07-FQM-02962.
\end{theacknowledgments}

\end{document}